\def\deg{\ensuremath{^\circ}}
\def\la{\mathrel{\mathchoice {\vcenter{\offinterlineskip\halign{\hfil
$\displaystyle##$\hfil\cr<\cr\sim\cr}}}
{\vcenter{\offinterlineskip\halign{\hfil$\textstyle##$\hfil\cr
<\cr\sim\cr}}}
{\vcenter{\offinterlineskip\halign{\hfil$\scriptstyle##$\hfil\cr
<\cr\sim\cr}}}
{\vcenter{\offinterlineskip\halign{\hfil$\scriptscriptstyle##$\hfil\cr
<\cr\sim\cr}}}}}
\def\ga{\mathrel{\mathchoice {\vcenter{\offinterlineskip\halign{\hfil
$\displaystyle##$\hfil\cr>\cr\sim\cr}}}
{\vcenter{\offinterlineskip\halign{\hfil$\textstyle##$\hfil\cr
>\cr\sim\cr}}}
{\vcenter{\offinterlineskip\halign{\hfil$\scriptstyle##$\hfil\cr
>\cr\sim\cr}}}
{\vcenter{\offinterlineskip\halign{\hfil$\scriptscriptstyle##$\hfil\cr
>\cr\sim\cr}}}}}

      [1999/12/01 v1.4c Il Nuovo Cimento]
\documentclass{cimento}
\usepackage{graphicx}  
\title{Studies of Gamma Ray Sources with the Fermi Large Area Telescope}
\author{J.~Kn\"odlseder\from{ins:cesr},
              on behalf of the Fermi Large Area Telescope Collaboration}
\instlist{\inst{ins:cesr} CESR, CNRS/UPS, - 9, avenue du Colonel Roche, 
                                       31028 Toulouse, France}
\PACSes{\PACSit{95.85.Pw}{Gamma rays}
\PACSit{98.70.Sa}{Cosmic rays}}
\begin{document}

\maketitle

\begin{abstract}
With its excellent sensitivity, large field of view, broad energy coverage, and good per-photon
angular resolution, the Large Area Telescope aboard the \textit{Fermi Gamma-ray Space Telescope}
satellite provides us with an unprecedented view of the high-energy Universe, revealing
a large diversity of cosmic particle accelerators that are active at various scales.
We present in this paper a selection of science highlights of the \textit{Fermi} mission, with
particular emphasis on results that are relevant for cosmic-ray physics.
We cover observations of supernova remnants and studies of interstellar gamma-ray
emission, reaching from the vicinity of the solar system out to the more distant starburst
galaxies.
\end{abstract}

\section{Introduction}

The \textit{Fermi Gamma-ray Space Telescope} has been successfully launched on 2008 June
11, and since 2008 August it routinely surveys the sky with the Large Area Telescope (LAT).
Compared to its predecessor, the EGRET telescope that operated aboard \textit{CGRO}
from 1991 to 2000, the LAT brings a sensitivity improvement of more than a factor of 10,
provides a wide field of view ($2.4$ sr at 1~GeV), and covers a wide energy range
from 20~MeV to $>300$~GeV \cite{atwood09}.
In its regular surveying mode, the entire sky is observed every 3 hours, providing information
on flux variability for any source in the sky.

We present in this paper a selection of the science highlights of the \textit{Fermi} mission, with
particular emphasis on results that are relevant for cosmic-ray physics.
While we focus here on observations of gamma rays, an accompanying paper will present
results on direct cosmic-ray measurements \textit{Fermi} (Latronico, these proceedings).
We start with an overview of the types of (steady) gamma-ray sources that are observed by
\textit{Fermi}/LAT, and summarize results that were obtained on observations of Galactic
supernova remnants.
The remainder of the paper is dedicated to the observation of interstellar gamma-ray emission,
and results are presented by gradually moving away from the local interstellar medium to the extragalactic space.

\section{Gamma-ray sources}

\subsection{Source Catalogue}

\begin{figure}
\centering
\includegraphics{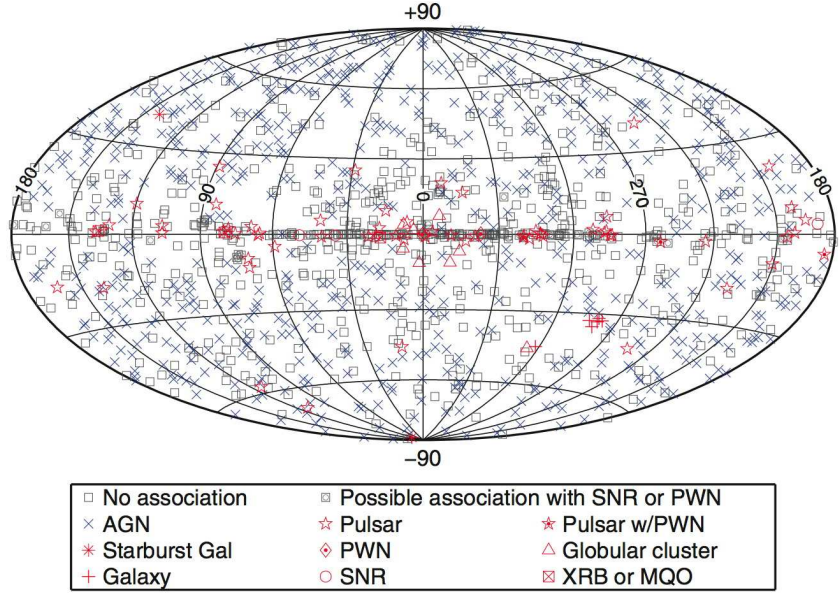}
\caption{1451 1FGL catalogue sources, showing locations on the sky (in Galactic coordinates
with Aitoff projection) and associated source class, coded according to the legend. The colour
is chosen simply to enhance the visibility of the associated and non-blazar sources (from 
\cite{abdo10a}).}
\label{fig:sourcecat}
\end{figure}

The combination of deep and fairly uniform exposure, good per-photon angular resolution,
and stable response of the LAT has made for the most sensitive, best-resolved survey of
the sky to date in the 100 MeV to 100 GeV energy range.
During the first 11 months of operations, 1451 gamma-ray sources have been
significantly detected in this energy range by the LAT, making up the First \textit{Fermi}-LAT
catalogue (1FGL) \cite{abdo10a} that provides a significant enhancement over the catalogue
of 271 sources that have been by detected the EGRET telescope over its entire mission 
lifetime \cite{hartman99}.

About 56\% of the 1FGL sources have been associated to counterparts at other wavelengths
based on positional coincidence at the $80\%$ confidence level (cf.~Fig.~\ref{fig:sourcecat}).
The large majority of these associations are with Blazars, i.e.~active galactic nuclei (AGN)
that presumably host supermassive black holes creating relativistic jets that are pointing in 
the general direction of the Earth and that are the sites of particle acceleration.
The second most important source class are pulsars (56 sources) which are firmly identified
by the high-confidence detection of periodicity in the arrival times of the gamma-ray
photons that is caused by the rotation of the neutron star.
While most of the pulsars are young and energetic, an increasingly large number of
millisecond pulsars has also been detected by the LAT \cite{abdo09a}.
Populations of millisecond pulsars are also believed to account for the gamma-ray
emission that is seen towards globular clusters \cite{abdo09b} of which 8 are
associated with gamma-ray sources in the 1FGL catalogue.
Young and energetic pulsars are often associated to pulsar wind nebulae (PWN)
and the remnants of their natal supernova explosions (SNR), and consequently there
is some ambiguity in associating gamma-ray sources to these 3 source classes.
Excluding 1FGL sources that are associated with pulsars, we find 6 1FGL sources
associated to PWN and 41 1FGL sources associated to SNRs.
Dedicated follow-up studies that investigate the spectral energy distributions and
spatial morphologies of the sources may help to clarify the underlying natures of
these sources (cf.~Section~\ref{sec:snr}).
Finally, 3 1FGL sources are firmly identified as high-mass X-ray binaries based on their
orbital variability 
(LS~I$+61\deg303$ \cite{abdo09c}, LS~5039 \cite{abdo09d} and Cyg~X-3 \cite{abdo09e}),
and 2 1FGL sources are associated to the starburst galaxies M~82 and NGC~253 
\cite{abdo10b}.
Furthermore, gamma-ray emission is detected from the LMC \cite{abdo10c} and SMC
\cite{abdo10d} and several 1FGL sources are associated to local emission maxima in these 
dwarf galaxies.

\subsection{Supernova remnants}
\label{sec:snr}

Supernova remnants have long been considered the primary candidates for the
origin of Galactic cosmic rays (CRs).
Specifically, diffusive shock acceleration \cite{bell78,blandford78} is widely accepted
as the mechanism by which charged particles can be accelerated to very high
energies at collisionless shocks driven by supernova explosions.
However, it has not yet been confirmed whether strong shocks in SNRs are indeed
capable of efficiently transferring kinetic energy into the acceleration of CR
ions, and the definite proof for cosmic-ray acceleration in SNRs is still missing.

Gamma-ray observations may probe for ion acceleration in SNRs by revealing
the characteristic decay signature from $\pi^0$ mesons that are produced by collisions 
between relativistic nuclei with ambient gas.
Recent ground-based gamma-ray observations in the TeV domain have revealed
several spatially resolved young SNRs (age $\sim1$ kyr) in our Galaxy, showing a 
morphology that correlates well with that observed in non-thermal X rays.
The TeV observations, however, cover only the high-energy part of the source
spectra, and thus discriminate only poorly between leptonic (inverse Compton 
and/or Bremsstrahlung) and hadronic ($\pi^0$ decay) emission scenarios.
Clearly, an improved low-energy coverage of these sources is needed to
better constrain the underlying emission mechanism \cite{funk07}.

\begin{figure}
\centering
\includegraphics[scale=2.0]{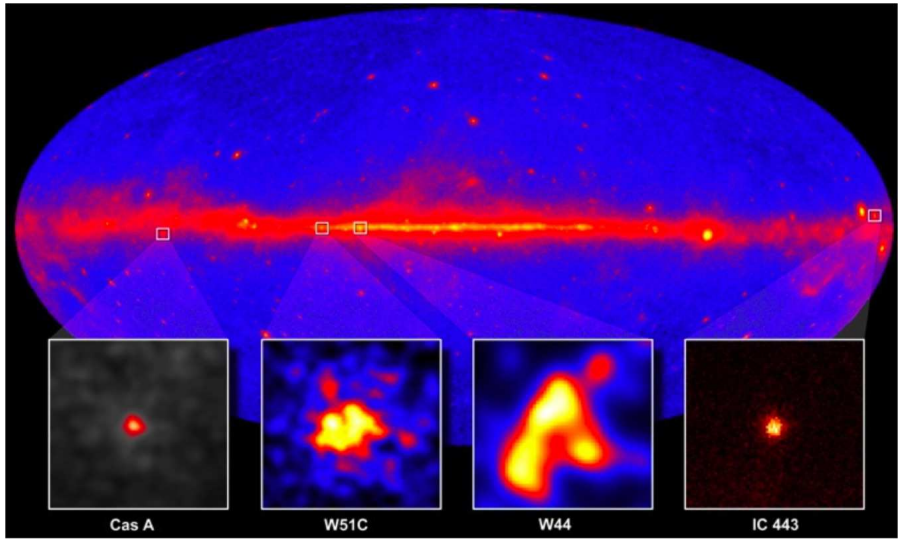}
\caption{\textit{Fermi}/LAT allsky map of the gamma-ray sky with enlarged images of four SNRs
superimposed.}
\label{fig:snr}
\end{figure}

\textit{Fermi} has now provided this low-energy coverage for several galactic SNRs
(cf.~Fig.~\ref{fig:snr}).
Among those are Cas~A \cite{abdo10e} and RX~J1713.7-3946 that both have been
detected also at TeV gamma rays.
Both objects are young SNRs that exhibit broad-band emission spectra ranging
from $\la1$~GeV up to $\sim10$~TeV and beyond.
The spectra of these sources impose so far only little constraints on the underlying
emission mechanism and are satisfactorily modelled by either leptonic or hadronic
emission models.
Interestingly, regardless of the origin(s) of the observed gamma rays, the total
amount of CRs accelerated in Cas~A constitutes only a minor fraction ($\le2\%$)
of the total kinetic energy of the supernova \cite{abdo10e}.

\begin{figure}
\centering
\includegraphics{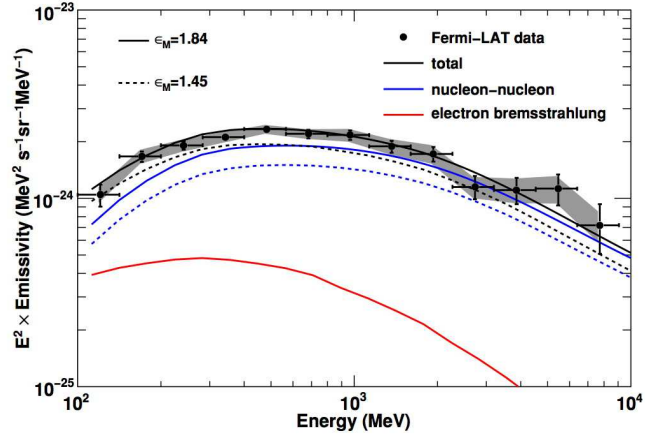}
\caption{Differential gamma-ray emissivity from the local atomic hydrogen gas compared with
the calculated gamma-ray production (from \cite{abdo09g}). 
}
\label{fig:lism}
\end{figure}

A second class of SNRs detected by \textit{Fermi} consists of mid-aged ($3-50$~kyr)
remnants that generally are known to be interacting with molecular clouds that
might act as target material for $\pi^0$ production.
These SNRs (W51C \cite{abdo09f}, W44 \cite{abdo10f}, IC~433 \cite{abdo10g},
W28 \cite{abdo10h}) are all characterized by spatially extended emission in the 
GeV domain with spectral breaks near $\sim1$~GeV and a spectral steepening
towards the TeV domain.
Consequently, many of these SNRs are only barely detected at TeV energies with 
current Cerenkov telescopes.
The gamma-ray spectra of these objects can be fitted with either leptonic and hadronic
models, yet in general, leptonic models require rather extreme conditions
to be met to explain the observations
(such as unusually large ratios of injected electrons to protons, strong magnetic
fields, ad-hoc breaks in the particle spectra, excessively large ambient photon 
densities or electron energy contents; e.g.~\cite{abdo09f,abdo10f,abdo10g,abdo10h}).
Consequently, hadronic models provide a more plausible explanation of the
observed emissions.
The increasing amount of SNRs that are detected by \textit{Fermi} together with a
continuously growing database at TeV energies opens up the possibility to study now 
how particle acceleration responds to environmental effects such as shock
propagation in dense clouds and how accelerated particles are released into
interstellar space.

\section{Interstellar Gamma-Ray Emission}

\subsection{Galactic cosmic rays}

Once accelerated, CRs diffuse away from their acceleration sites into the interstellar space
of our Galaxy where in encounters with the interstellar gas and radiation fields they
produce a diffuse glow of gamma rays through $\pi^0$ decay, inverse Compton scattering
and non-thermal Bremsstrahlung processes.
This diffuse Galactic glow is in fact the first source of high-energy gamma rays
that was discovered by observations with the OSO-3 satellite in 1968 \cite{clark68}.
Similar to observations of SNRs, the study of the diffuse Galactic gamma-ray emission
provides important insights into CR acceleration and propagation within our Galaxy.

Figure \ref{fig:lism} shows the differential gamma-ray emissivity of hydrogen in the local 
interstellar medium (within 1 kpc of the solar system) as determined by \textit{Fermi}
from observations of diffuse gamma-ray emission at high Galactic latitudes \cite{abdo09g}.
The differential emissivity spectrum agrees remarkably well with calculations based on
CR spectra that are consistent with those measured directly at Earth, at the $10\%$ level.
This indicates that the CR nuclei spectra within 1 kpc from the solar system are comparable
to those measured near Earth.

Going further away from the Sun, observations of the gamma-ray emission towards the
Cassiopeia and Cepheus constellations allow studying CR density variations in the outer
Galaxy, covering Galactocentric distances from $\sim9$ kpc to $\sim20$ kpc.
A recent \textit{Fermi} study of this region revealed that the gamma-ray emissivity
spectrum of the gas in the nearby Gould Belt (within 300 pc from the solar system) is 
consistent with expectations based on locally measured CR spectra \cite{abdo10i}.
The gamma-ray emissivity decreases from the Gould Belt to the distant Perseus arm,
but the measured gradient is flatter than expectations based on current estimates of the 
distribution of sources of CRs in the Milky Way and of CR propagation parameters.
In addition, the observations present evidence in the Gould Belt for so called dark gas,
which is gas that is not properly traced by radio and microwave surveys, and of which the
mass amounts to $\sim50\%$ of the CO-traced molecular gas mass.

\begin{figure}
\centering
\includegraphics[scale=0.4]{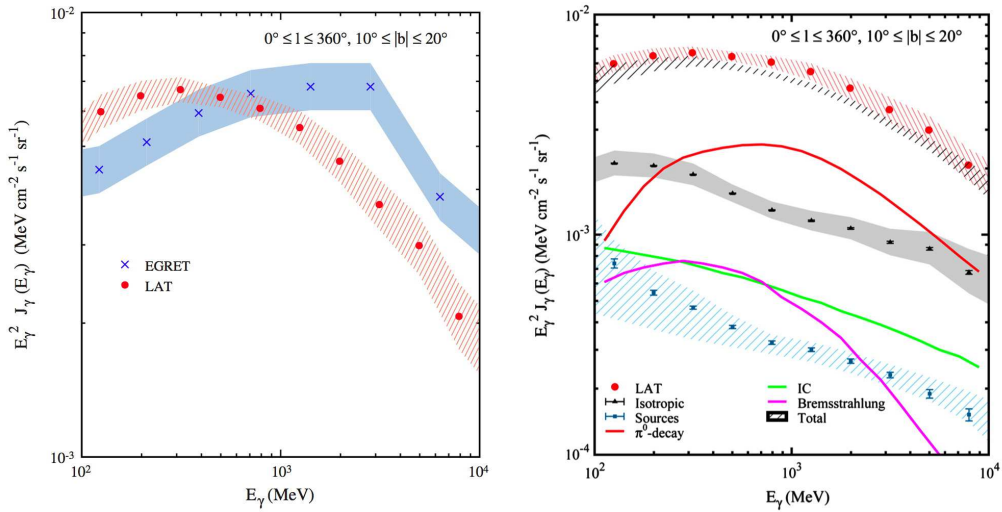}
\caption{Diffuse emission intensity averaged over all galactic longitudes for latitude
range $10\deg \le |b| \le 20\deg$. The left plot shows the comparison of LAT data
(red) to EGRET data (blue), the right plot shows the comparison of LAT data with a model
of Galactic diffuse emission (from \cite{abdo09h}).}
\label{fig:gevexcess}
\end{figure}

In the late nineties, measurements with the EGRET telescope 
indicated a global excess of diffuse emission $\ga1$~GeV relative to that
expected from conventional diffuse galactic emission models \cite{hunter97} which led
to speculations about a possible dark matter origin of this so-called ``GeV excess''
\cite{deboer05}.
\textit{Fermi} has measured the diffuse gamma-ray emission with improved sensitivity 
and resolution with respect to EGRET \cite{abdo09h}.
Figure \ref{fig:gevexcess} compares the LAT data (red) to the earlier EGRET data (blue)
which reveals a significant discrepancy between both measurements.
In particular, the LAT data do not show the excess reported by EGRET and are in fact
well reproduced by a diffuse gamma-ray emission model that is consistent with local
CR spectra.
The knowledge about the LAT instrument response function comes from detailed
simulations that were validated with beam tests of calibration units, and from post-launch
refinements based on the actual particle background, and are considered as accurate
\cite{abdo09h}.
It thus is plausible to attribute the ``GeV excess'' to an instrumental artefact of the earlier 
EGRET measurements.

\subsection{Mapping cosmic-ray acceleration in the LMC}

\begin{figure}
\centering
\includegraphics[scale=0.37]{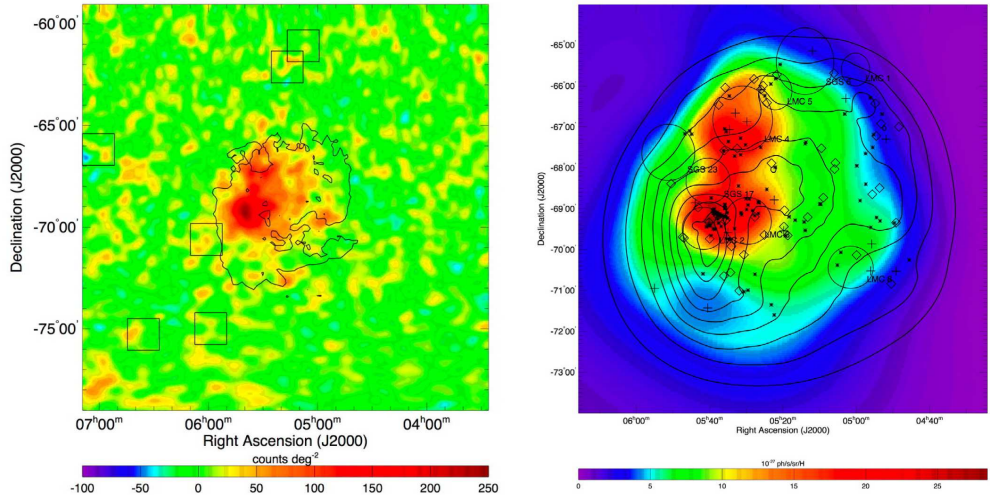}
\caption{Left panel: Gaussian kernel ($\sigma=0.2\deg$) smoothed background subtracted counts
map of the LMC region for the energy range 200 MeV - 20 GeV. 
Right panel: Adaptively smoothed integrated $>100$ MeV emissivity map of the LMC revealing
the cosmic-ray density distribution within the LMC. The contours indicate the gas density
smoothed with the LAT instrumental point spread function (from \cite{abdo10k}).}
\label{fig:lmc}
\end{figure}

Nearby galaxies have some advantages as targets for studies of CR physics as they are
viewed from outside, and so line of sight confusion, which complicates studies of gamma-ray
emission from our own Galaxy, is diminished.
The Large Magellanic Cloud (LMC) is an excellent target for studying the link between
CR acceleration and gamma-ray emission since the galaxy is nearby ($\sim50$~kpc),
can be easily resolved (angular extent $\sim8\deg$), and is seen almost face-on.
The LMC has been initially detected by the EGRET telescope \cite{sreekumar92}, 
but \textit{Fermi} now provides the instrumental capabilities to perform a detailed study
of the galaxy.

The LMC is clearly detected with the LAT ($\ga33\sigma$) and for the first time the
emission is spatially well resolved in gamma rays \cite{abdo10k}.
The observations (cf.~Fig.~\ref{fig:lmc}) reveal the massive star forming region 30~Doradus 
as a bright source of gamma-ray emission, in addition to a fainter glow that spreads out 
over large areas of the galaxy.
Surprisingly, the observations reveal little correlation of the gamma-ray emission with
gas density, as it would have been expected if CRs propagate throughout the entire 
galaxy.
The gamma-ray emission correlates more with tracers of massive star forming
regions, supporting the idea that CRs are accelerated in these regions as a result of the
large amounts of kinetic energy that are input by the stellar winds and supernova
explosions of massive stars into the interstellar medium.

\subsection{Starburst galaxies}

\begin{figure}
\centering
\includegraphics[scale=1.0]{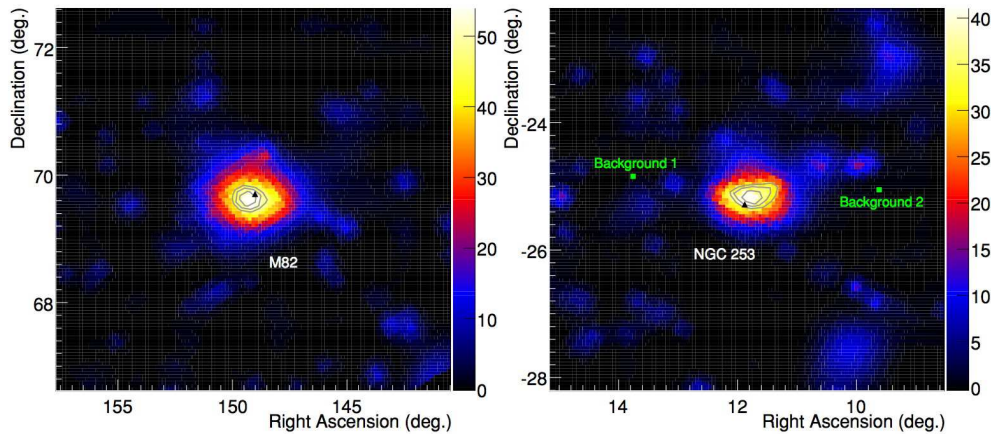}
\caption{Test statistic maps obtained from photons above 200 MeV showing the 
$6\deg$ by $6\deg$ large regions around M~82 and NGC~253.
Aside from the source associated with each galaxy, all other \textit{Fermi}-detected
sources (green squares) within a $10\deg$ radius of the best-fit position have been 
included in the background model as well as components describing the diffuse Galactic 
and isotropic gamma-ray emission. Black triangles denote the centres of
M~82 and NGC~253 at optical wavelengths (from \cite{abdo10l}).}
\label{fig:starbursts}
\end{figure}

Probing galactic cosmic-ray acceleration even well beyond the local group of galaxies
has now become possible thanks to \textit{Fermi}.
For the first time, steady GeV gamma-ray emission has been detected significantly by the
LAT from sources positionally coincident with locations of the starburst galaxies M~82 and 
NGC~253 \cite{abdo10l}.
Test statistic maps obtained with LAT for photons $\ge200$~MeV for regions around
both galaxies are shown in Fig.~\ref{fig:starbursts}.
Both starburst galaxies have also been detected at TeV energies by VERITAS \cite{acciari09}
and H.E.S.S. \cite{acero09}, and the emission is well explained by the interaction of 
CRs with local interstellar gas and radiation fields.
M~82 and NGC~253, though having less gas than the Milky Way, have factors of $2-4$
greater gamma-ray luminosity, suggesting a connection between active star formation and
enhanced CR energy densities in star-forming galaxies.
In particular, the H.E.S.S. observations of NGC~253 localise the gamma-ray emission towards
the starbursting core of the galaxy, which is very much like the situation in the LMC where the 
gamma-ray emission is brightest towards 30 Doradus, a region that is considered
as a ``mini-starburst'' \cite{redman03}.
Star-forming and starburst galaxies are thus a new class of prominent gamma-ray
emitters in the Universe, and thanks to their large number, they have the potential to make
a significant, $\ga10\%$ contribution to the extragalactic gamma-ray background at
high-energy gamma rays \cite{abdo10l}.

\acknowledgments
The \textit{Fermi} LAT Collaboration acknowledges generous ongoing support
from a number of agencies and institutes that have supported both the
development and the operation of the LAT as well as scientific data analysis.
These include the National Aeronautics and Space Administration and the 
Department of Energy in the United States, the Commissariat \`a l'\'Energie Atomique
and the Centre National de la Recherche Scientifique / Institut National de Physique
Nucl\'eaire et de Physique des Particules in France, the Agenzia Spaziale Italiana
and the Istituto Nazionale di Fisica Nucleare in Italy, the Ministry of Education,
Culture, Sports, Science and Technology (MEXT), High Energy Accelerator Research
Organization (KEK) and Japan Aerospace Exploration Agency (JAXA) in Japan, and
the K.~A.~Wallenberg Foundation, the Swedish Research Council and the
Swedish National Space Board in Sweden.

Additional support for science analysis during the operations phase is gratefully
acknowledged from the Istituto Nazionale di Astrofisica in Italy and the 
Centre National d'\'Etudes Spatiales in France.



\begin{thebibliography}{0}
\bibitem{atwood09} \BY{Atwood~W.B. et al.}
  \IN{ApJ}{697}{2009}{1071}.
\bibitem{abdo10a} \BY{Abdo~A.A. et al.}
  \IN{ApJS}{188}{2010a}{405}.
\bibitem{hartman99} \BY{Hartman~R.C. et al.}
  \IN{ApJS}{123}{1999}{79}.
\bibitem{abdo09a} \BY{Abdo~A.A. et al.}
  \IN{Science}{325}{2009a}{848}.
\bibitem{abdo09b} \BY{Abdo~A.A. et al.}
  \IN{Science}{325}{2009b}{845}.
\bibitem{abdo09c} \BY{Abdo~A.A. et al.}
  \IN{ApJ}{701}{2009c}{123}.
\bibitem{abdo09d} \BY{Abdo~A.A. et al.}
  \IN{ApJ}{706}{2009d}{L56}.
\bibitem{abdo09e} \BY{Abdo~A.A. et al.}
  \IN{Science}{326}{2009e}{1512}.
\bibitem{abdo10b} \BY{Abdo~A.A. et al.}
  \IN{ApJ}{709}{2010b}{L152}.
\bibitem{abdo10c} \BY{Abdo~A.A. et al.}
  \IN{A\&A}{512}{2010c}{A7}.
\bibitem{abdo10d} \BY{Abdo~A.A. et al.}
  \IN{A\&A}{}{2010d}{in press}.
\bibitem{bell78} \BY{Bell~A.R.}
  \IN{MNRAS}{}{1978}{147}.
\bibitem{blandford78} \BY{Blandford~R.D. \& Ostriker~J.A.}
  \IN{ApJ}{221}{1978}{L29}.
\bibitem{funk07} \BY{Funk~S.}
  \IN{30th ICRC Conference}{}{2007}{astro-ph/0709-3127v1}.
\bibitem{abdo10e} \BY{Abdo~A.A. et al.}   
  \IN{ApJ}{710}{2010e}{L92}.
\bibitem{abdo09f} \BY{Abdo~A.A. et al.}  
  \IN{ApJ}{706}{2009f}{L1}.
\bibitem{abdo10f} \BY{Abdo~A.A. et al.}   
  \IN{Science}{327}{2010f}{1103}.
\bibitem{abdo10g} \BY{Abdo~A.A. et al.}   
  \IN{ApJ}{712}{2010g}{459}.
\bibitem{abdo10h} \BY{Abdo~A.A. et al.}   
  \IN{ApJ}{}{2010h}{in press}.
\bibitem{clark68} \BY{Clark~G.W. et al.}
  \IN{ApJ}{153}{1968}{203}.
\bibitem{abdo09g} \BY{Abdo~A.A. et al.}  
  \IN{ApJ}{703}{2009g}{1249}.
\bibitem{abdo10i} \BY{Abdo~A.A. et al.}   
  \IN{ApJ}{710}{2010i}{133}.
\bibitem{hunter97} \BY{Hunter~S.D. et al.}
  \IN{ApJ}{481}{1997}{205}.
\bibitem{deboer05} \BY{de Boer~W. et al.}
  \IN{A\&A}{444}{2005}{51}.
\bibitem{abdo09h} \BY{Abdo~A.A. et al.}  
  \IN{PRL}{103}{2009h}{251101}.
\bibitem{sreekumar92} \BY{Sreekumar et al.}
  \IN{ApJ}{400}{1992}{L67}.
\bibitem{abdo10k} \BY{Abdo~A.A. et al.}   
  \IN{A\&A}{512}{2010k}{A7}.
\bibitem{abdo10l} \BY{Abdo~A.A. et al.}   
  \IN{ApJ}{709}{2010l}{L152}.
\bibitem{acciari09} \BY{Acciari~V.A. et al.}
  \IN{Nature}{462}{2009}{770}.
\bibitem{acero09} \BY{Acero~F. et al.}
  \IN{Science}{326}{2009}{1080}.
\bibitem{redman03} \BY{Redman~M.P. et al.}
  \IN{MNRAS}{344}{2003}{741}.
\end{thebibliography}
\end{document}